\documentclass[11pt,a4paper]{article}
\usepackage{jheppub}
\usepackage{amstext}

\title{Properties of 125 GeV Higgs boson\\
 in non-decoupling MSSM scenarios}
\author[a]{Kaoru Hagiwara,}
\author[b]{Jae Sik Lee,}
\author[a]{Junya Nakamura}

\affiliation[a]{KEK Theory Center and Sokendai, Tsukuba, Ibaraki 305-0801, JAPAN}
\affiliation[b]{Department of Physics, National Tsing Hua University, Hsinchu, Taiwan
300}

\emailAdd{kaoru.hagiwara@kek.jp}
\emailAdd{jslee@phys.nthu.edu.tw}
\emailAdd{junnaka@post.kek.jp}

\abstract{Tantalizing hints of the Higgs boson of mass around 125 GeV have been reported at the LHC. We explore the MSSM parameter space in which the 125 GeV state is identified as the heavier of the CP even Higgs bosons, and study two scenarios where the two photon production rate can be significantly larger than the standard model (SM). In one scenario, $\Gamma(H\to \gamma\gamma)$ is enhanced by a light stau contribution, while the $WW^{\ast}$ ($ZZ^{\ast}$) rate stays around the SM rate. In the other scenario, $\Gamma(H\to b\bar{b})$ is suppressed and not only the $\gamma\gamma$ but also the $WW^{\ast}$ ($ZZ^{\ast}$) rates should be enhanced. The $\tau\bar{\tau}$ rate can be significantly larger or smaller than the SM rate in both scenarios. Other common features of the scenarios include top quark decays into charged Higgs boson, single and pair production of all Higgs bosons in $e^+e^-$ collisions at $\sqrt{s}\lesssim 300$ GeV.}


\begin{document}
\maketitle
\flushbottom

\section{Introduction}
Latest results from the Higgs boson search by the ATLAS \cite{higgsatlas} and the CMS \cite{higgscms} collaborations show an excess of events around the mass region of 125 GeV. The main search channel is the two photons decay mode of the Higgs boson, for which both experiments reported the rate higher than the standard model (SM) prediction. There are hints of the $ZZ^*$ decay mode with less significance, while no hints have been reported for the $\tau\bar{\tau}$ mode. We expect that the data from the current 8 TeV run will make clear the properties of the Higgs boson candidate.

The Higgs sector in the minimal supersymmetric standard model (MSSM) has five physical mass eigenstates, two CP even and one CP odd neutral scalar bosons, if CP is conserved in the Higgs sector, and one pair of charged scalar bosons \cite{susy}. The observed $\gamma\gamma$ resonance at 125 GeV can be one of the three neutral Higgs bosons. Among them, the CP odd state ($A$) cannot give the $\gamma\gamma$ rate greater than that of the SM Higgs bosons, mainly because it lacks the W boson loop contribution to the $\gamma\gamma$ decay \cite{wloop}. Among the two CP even Higgs bosons, both the light (h) and heavy (H) mass eigenstates can be 125 GeV and can have enhanced $\gamma\gamma$ rate. MSSM scenarios where the lighter of the CP even Higgs boson is identified as the 125 GeV state are discussed in refs. \cite{refh1,refh2,refh3,refh4,refh5,refh6,refh7,refh8,refh9,refh10,refh11,refh12*,refh13,refh14,refh15,refh16,refh17}, and the possibility of the 125 GeV state as the heavier of the CP even Higgs bosons is discussed in ref. \cite{refh12*}. The former scenario contains the so-called decoupling region where all the other Higgs bosons ($H$, $A$, $H^{\pm}$) are significantly heavier than the lighter CP even state $h$, whose properties resembles the SM Higgs boson. On the other hand, in the latter scenario where the heavier of the CP even state $H$ has the mass 125 GeV, not only the mass of the lighter CP even state $h$ but also those of the CP odd state $A$ and the charged Higgs boson $H^{\pm}$ are bounded from above.

In this study, we study carefully the consequences of this non-decoupling scenario of MSSM where the 125 GeV state is the heavier of the CP even Higgs bosons, $H$. In particular, we study two sub-scenarios where the two photon production rate can be larger than the SM. In one scenario, the $H\to\gamma\gamma$ amplitude is enhanced by a light stau loop which interferes constructively with the main W boson loop, while the $WW^*$ ($ZZ^*$) rate is around the SM prediction. In another scenario, the $\gamma\gamma$ rate is enhanced by suppressing the dominant partial decay width $\Gamma(H\to b\bar{b})$, and not only $\gamma\gamma$ but also $WW^{\ast}$ ($ZZ^{\ast}$) production rate can be large. In both scenarios, the $\tau\bar{\tau}$ rate can be significantly larger or smaller than in the SM. Prediction for the mass spectra of the other Higgs bosons is also examined.

The enhancement of the two photon production rate due to a light stau in the decoupling region has been studied in refs. \cite{refh8,refh16}. We show in this report that the same mechanism works in the non-decoupling region as well. The suppression of $\Gamma(H\to b\bar{b})$ in the non-decoupling region has been studied in ref. \cite{refh12*}. We study not only the $\gamma\gamma$ and $WW^{\ast}$ ($ZZ^{\ast}$) rates but also the $\tau\bar{\tau}$ rate in detail.


\section{Higgs sector in MSSM}
In this section, we briefly review the mass spectrum of the Higgs bosons in MSSM. In our scenarios where the two photon production rate of the heavier CP even state $H$ is higher than that of the SM, relatively large Higgs couplings to the weak bosons are necessary, since the main contribution to the $H \to \gamma\gamma$ amplitude comes from the W boson loop \cite{wloop}. Hence, $H$ must be a SM-like Higgs boson.

The MSSM Higgs sector consists of two $SU(2)_L$ doublets, $\phi_u$ and $\phi_d$ which give masses to up type fermions and down type fermions, respectively \cite{susy}.  When the electroweak symmetry is spontaneously broken, MSSM gives five physical mass eigenstates, two CP even scalar bosons $h$ and $H$, one CP odd scalar boson $A$, and one pair of charged scalar bosons $H^{\pm}$. The two CP even scalar bosons are mixed states of the real components of the two Higgs doublets, 
\begin{align}
\begin{pmatrix}
h\\
H
\end{pmatrix}
=
\begin{pmatrix}
\cos\alpha & -\sin\alpha\\
\sin\alpha & \cos\alpha
\end{pmatrix}
\begin{pmatrix}
H_u^0\\
H_d^0
\end{pmatrix},\label{alpha}
\end{align}
where we define $h$ and $H$ as the lighter and the heavier of the two CP even scalar bosons, respectively, whereas the current basis states $H_u^0$ and $H_d^0$ are defined as in
\begin{align}
Re (\phi_u^0)=\frac{v_u+H_u^0}{\sqrt{2}},\ \ Re(\phi_d^0)=\frac{v_d+H_d^0}{\sqrt{2}}.
\end{align}
Upon the convention that the above vacuum expectation values are written as $v_u=v\sin\beta$ and $v_d=v\cos\beta$ with $v$ $(\simeq 245)$ GeV being the vacuum expectation value of the SM Higgs doublet, we can introduce another base
\begin{align}
\begin{pmatrix}
H_u^0\\
H_d^0
\end{pmatrix}
=
\begin{pmatrix}
\sin\beta & -\cos\beta\\
\cos\beta & \sin\beta
\end{pmatrix}
\begin{pmatrix}
H_{SM}\\
H_{\bot}
\end{pmatrix},\label{beta}
\end{align}
where $H_{SM}$ is a state whose couplings to the weak bosons are the same as those of the SM Higgs boson, and $H_{\bot}$ is its orthogonal state which has no coupling to the weak bosons. From eqs. (\ref{alpha}, \ref{beta}), we have
\begin{subequations}
\begin{align}
h&=-\sin{(\alpha-\beta)}H_{SM}-\cos{(\alpha-\beta)}H_{\bot},\\
H&=\cos{(\alpha-\beta)}H_{SM}-\sin{(\alpha-\beta)}H_{\bot}.\label{H}
\end{align}
\end{subequations}
The masses and the eigenstates of the CP even Higgs bosons in the MSSM are determined by diagonalizing the symmetric mass-squared matrix in the space of $(H_u^0, H_d^0)^T$, 
\begin{align}
\begin{pmatrix}
M_{uu}^2 & M_{ud}^2\\
M_{ud}^2 & M_{dd}^2
\end{pmatrix}, \label{massmatrix}
\end{align}
whose elements can be approximated as \cite{Higgsmass1, Higgsmass2}
\begin{subequations}\label{masselement}
\begin{align}
M_{uu}^2&\sim M_Z^2\left(1-\frac{3}{8\pi^2}Y_t^2\ln{\frac{M_{susy}^2}{M_t^2}}\right)+\frac{3v^2}{8\pi^2}Y_t^4\left[\ln{\frac{M_{susy}^2}{M_t^2}}+\bar{A}_t^2\left(1-\frac{\bar{A}_t^2}{12}\right)\right]
-\frac{3v^2}{96\pi^2}Y_b^4\bar{\mu}^4,\label{Muu} \\
M_{dd}^2&\sim M_A^2-\frac{v^2}{32\pi^2}Y_t^4\bar{\mu}^2\bar{A}_t^2-\frac{v^2}{32\pi^2}Y_b^4\bar{\mu}^2\bar{A}_b^2,\label{Mdd} \\ 
M_{ud}^2&\sim -\cos\beta\left[M_A^2+M_Z^2+\frac{v^2}{16\pi^2}Y_t^4\bar{\mu}^2(\bar{A}_t^2-3)+\frac{v^2}{16\pi^2}Y_b^4\bar{\mu}^2(\bar{A}_b^2-3)\right] \nonumber \\
&+\frac{v^2}{32\pi^2}Y_t^4\bar{\mu}\bar{A}_t\left(\bar{A}_t^2-6\right)+\frac{v^2}{32\pi^2}Y_b^4\bar{\mu}^3\bar{A}_b,\label{Mud}
\end{align}
\end{subequations}
where only the leading terms for large $\tan\beta$ ($\tan\beta \gg 1$) are kept, since large value of $\tan\beta$ is necessary to have a SM-like Higgs boson as heavy as $125$ GeV. $Y_t$ and $Y_b$ are, respectively, the top and bottom Yukawa couplings in the MSSM. The soft SUSY breaking $A_f$ terms and the Higgsino mass $\mu$ are made dimensionless as $\bar{A}_t=A_t/M_{susy}$, $\bar{A}_b=A_b/M_{susy}$, $\bar{\mu}_t=\mu/M_{susy}$ with
\begin{align}
M_{susy}^2=\frac{M^2_{\tilde{t}_1}+M^2_{\tilde{t}_2}}{2},\label{msusy}
\end{align}
where $M_{\tilde{t}_1}$ and $M_{\tilde{t}_2}$ are masses of the stop mass eigenstates. Full analytic formulae of the mass matrix elements can be found in ref. \cite{Higgsmass1, Higgsmass2}. 
By diagonalizing the matrix eq. (\ref{massmatrix}), we obtain the masses of the two CP even Higgs bosons,
\begin{subequations}\label{mhandmH}
\begin{align}
M_h^2&=M_{uu}^2\cos^2{\alpha}+M_{dd}^2\sin^2{\alpha}-M_{ud}^2\sin{2\alpha}, \label{mh}\\ 
M_H^2&=M_{uu}^2\sin^2{\alpha}+M_{dd}^2\cos^2{\alpha}+M_{ud}^2\sin{2\alpha}, \label{mH}
\end{align}
\end{subequations}
with
\begin{subequations}
\begin{align}
\sin\alpha&=\frac{M_{ud}^2}{\sqrt{(M_H^2-M_{uu}^2)^2+(M_{ud}^2)^2}},\label{sina}\\
\cos\alpha&=\frac{M_H^2-M_{uu}^2}{\sqrt{(M_H^2-M_{uu}^2)^2+(M_{ud}^2)^2}},\label{cosa}
\end{align}
\end{subequations}
where we can choose the mixing angle $\alpha$ in the region $-\frac{\pi}{2} < \alpha < \frac{\pi}{2}$, since $M_H^2-M_{uu}^2 > 0$ ($\cos\alpha > 0$) is always satisfied. The region of $\alpha$ can further be separated depending on the sign of $\sin\alpha$, 
\begin{subequations}\label{sina}
\begin{align}
\mathrm{1.}&\ \sin\alpha < 0\ \ (M_{ud}^2 < 0),\ \  -\frac{\pi}{2} < \alpha < 0,\label{-sina}\\ 
\mathrm{2.}&\ \sin\alpha > 0\ \ (M_{ud}^2 > 0),\ \ \ \ \ 0 < \alpha < \frac{\pi}{2}.\label{+sina}
\end{align}
\end{subequations}
Here the region 2 ($\sin\alpha > 0$) takes place if the loop contribution dominates over the negative definite tree-level contribution in eq. (\ref{Mud}), which can happen, for example when both $\bar{\mu}\bar{A}_t > 0$ and $\bar{A}_t^2 > 6$ are satisfied for tiny $\cos\beta$ (large $\tan\beta$)
\footnote[1]{Large values of $\bar{A}_t^2$ tend to give color- and charge-breaking minima, resulting in the bound $\bar{A}_t^2 < 15$; see for example ref. \cite{instability}.}.

In the limit that the state $H$ has the SM-like couplings to the weak bosons,
we have 
\begin{align}
|\cos(\alpha-\beta)|\simeq 1 \label{condition01}
\end{align}
in eq. (\ref{H}). At large $\tan\beta$ ($\beta \simeq \pi/2$), this condition eq. (\ref{condition01}) selects two distinct regions, $\alpha-\beta\simeq 0$ ($\alpha\simeq \pi/2$) or $\alpha-\beta \simeq -\pi$ ($\alpha\simeq -\pi/2$). In both cases, we have $\cos\alpha \ll 1$, and the mass eigenstates in eq. (\ref{mhandmH}) are approximated by
\begin{subequations}\label{mh3mH3}
\begin{align}
M_h^2\simeq M_{dd}^2-2M_{ud}^2\cos{\alpha}+O(\cos^2{\alpha}), \label{mh3} \\
M_H^2\simeq M_{uu}^2+2M_{ud}^2\cos{\alpha}+O(\cos^2{\alpha}). \label{mH3} 
\end{align}
\end{subequations}
By neglecting small terms proportional to $\cos\alpha$, the condition that the SM-like state $H$ is heavier than the other state approximately implies
\begin{align}
M_{uu}^2-M_{dd}^2 \gtrsim 0,\label{eq81}
\end{align}
or from eq. (\ref{masselement})
\begin{align}
&M_Z^2+\frac{3}{8\pi^2}Y_t^2\ln{\frac{M_{susy}^2}{M_t^2}}(v^2Y_t^2-M_Z^2)\nonumber \\
&+\frac{3}{8\pi^2}v^2Y_t^4\bar{A}_t^2\left(1-\frac{\bar{A}_t^2-\bar{\mu}^2}{12}\right)
+\frac{1}{32\pi^2}v^2Y_b^4\bar{\mu}^2(\bar{A}_b^2-\bar{\mu}^2) \gtrsim M_A^2.\label{eq42}
\end{align}
Hence $M_A$ is bounded from above by the loop contribution to the Higgs potential.

It should also be noted from eqs. (\ref{Muu}, \ref{mH3}) that, in order to make $H$ a SM-like Higgs boson and as heavy as 125 GeV, large $M_{susy}$ and $\bar{A}_t^2\sim 6$ are necessary, and we explore the MSSM parameter region which satisfies these conditions in the following sections.


\section{Scenarios giving large two photon rate}
In our analysis, we consider the following Higgs production processes at the LHC,
\begin{subequations}\label{process}
\begin{align}
\mathrm{Gluon\ fusion\ \ } gg \to& \ \phi + X,\\
\mathrm{Weak\ boson\ fusion\ \ } qq \to& \ qq\phi + X,\\
\mathrm{Bottom\ quark\ annihilation\ \ } b\bar{b} \to& \ \phi + X,
\end{align}
\end{subequations}
where $\phi$ can be $h, H$ or $A$.
The SM Higgs production cross sections for the processes in eq. (\ref{process}) are calculated by using the programs HIGLU \cite{HIGLU}, HAWK \cite{hawk1, hawk2} and BBH@NNLO \cite{bbhnnlo}, respectively. The MSSM Higgs cross sections are obtained by scaling the corresponding SM Higgs cross sections with the ratio of the corresponding MSSM decay width over the SM one. The decay widths, couplings and mass spectra of the Higgs bosons and SUSY particles are calculated with an updated version of CPsuperH2.0 \cite{cpsuperh1, cpsuperh2, cpsuperh3} which includes the stau contribution to the Higgs boson masses. Although the SM cross section of the bottom quark annihilation process is quite small compared to the dominant gluon fusion process, it can be significant in some MSSM scenarios. 

We consider the following constraints from the collider experiments. For the stau and stop masses, we adopt the lower mass bounds \cite{rpp}
\begin{subequations}\label{allconst}
\begin{align}
\mathrm{Stau\ }M_{\tilde{\tau}} > 81.9 \mathrm{\ GeV},\label{const1}\\
\mathrm{Stop\ }M_{\tilde{t}} > 92.6 \mathrm{\ GeV}.\label{const2}
\end{align}
Upper bounds on the $e^+e^-$ annihilation cross sections
\begin{align}
\sigma\bigl(e^+e^-\to Zh(\to Zb\bar{b} \mathrm{\ and\ } Z\tau\bar{\tau})\bigr),\label{const3}\\
\sigma\bigl(e^+e^-\to Ah(\to b\bar{b}b\bar{b},\ b\bar{b}\tau\bar{\tau}  \mathrm{\ and\ } \tau\bar{\tau}\tau\bar{\tau})\bigr)\label{const4}
\end{align}
are taken from ref. \cite{lep}, and those on the cross sections at the LHC
\begin{align}
\sigma\bigl(pp\to h,A,H(\to \tau\bar{\tau})\bigr)\label{const5}
\end{align}
are taken from ref. \cite{higgstautau}. Upper bound on the branching fraction 
\begin{align}
B\bigl(t\to bH^+(\to b\bar{\tau}\nu_{\tau})\bigr)\label{const6}
\end{align}
\end{subequations}
is taken from ref. \cite{chargedhiggs}. Since all the physical Higgs bosons are relatively light in our non-decoupling scenario when $M_h < M_H \approx 125$ GeV, all the above constraints in eqs. (\ref{allconst}) are required to be satisfied in all the results presented below. In particular, significant portion of  very large $\tan\beta$ regions is excluded by the $h, A, H\to \tau\bar{\tau}$ and $t\to bH^+(\to \bar{\tau}\nu_{\tau})$ search limits (eqs. (\ref{const5}, \ref{const6})).

We define the ratio of a production rate at the LHC as
\begin{align}
R_{AB}=\frac{\sigma(pp\to H)B(H\to AB)}{\sigma(pp\to H)^{SM}B(H\to AB)^{SM}},\label{Rab}
\end{align}
which gives the $H\to AB$ production rate normalized to the SM prediction. Although we calculate the Higgs boson cross section $\sigma(pp\to H+X)$ at $\sqrt{s}=7$ TeV in this study, the ratio $R_{AB}$ should not change significantly even for $\sqrt{s}=8$ TeV, since the dominance of the gluon fusion process remains to be valid. Hence, our results can also be applied for future results of $\sqrt{s}=8$ TeV. Since the gluon fusion process dominates over the other production processes and the total decay width is dominated by $\Gamma(H\to b\bar{b})$ for the heavy CP even state $H$ with mass around 125 GeV, $R_{AB}$ may be approximately written by using the partial decay widths,
\begin{align}
R_{AB}\simeq \left(\frac{\Gamma(H\to gg)}{\Gamma(H\to gg)^{SM}}\right) \cdot \left(\frac{\Gamma(H\to b\bar{b})}{\Gamma(H\to b\bar{b})^{SM}}\right)^{-1} \cdot \left(\frac{\Gamma(H\to AB)}{\Gamma(H\to AB)^{SM}}\right).\label{Rab2}
\end{align}
By introducing a short hand notation
\begin{align}
r_{ab}=\frac{\Gamma(H\to ab)}{\Gamma(H\to ab)^{SM}},\label{shortnote}
\end{align}
for a partial width normalized to the corresponding SM value, the production rate $R_{AB}$ of eq. (\ref{Rab2}) can be expressed as
\begin{align}
R_{AB}\simeq r_{gg}\cdot(r_{b\bar{b}})^{-1}\cdot r_{AB}.\label{Rab3}
\end{align}
In this study, we examine $R_{\gamma\gamma}$, $R_{VV}\ (V=W,\ Z)$ and $R_{\tau\bar{\tau}}$,
\begin{subequations}
\begin{align}
R_{\gamma\gamma}&\simeq r_{gg}\cdot(r_{b\bar{b}})^{-1}\cdot r_{\gamma\gamma} ,\label{Rpp}\\
R_{VV}&\simeq r_{gg}\cdot(r_{b\bar{b}})^{-1}\cdot r_{VV},\label{Rzz}\\
R_{\tau\bar{\tau}}&\simeq r_{gg}\cdot(r_{b\bar{b}})^{-1}\cdot r_{\tau\bar{\tau}},\label{Rtautau}
\end{align}
\end{subequations}
and identify two scenarios where the following two conditions are satisfied for the heavier CP even Higgs boson in the MSSM,
\begin{subequations}
\begin{align}
123 < &M_H < 127\ \mathrm{GeV},\\
1 < &R_{\gamma\gamma} < 3.
\end{align}
\end{subequations}
Specifically, they are
\begin{subequations}
\begin{align}
\mathrm{Light\ stau\ scenario}:&\ r_{\gamma\gamma} > 1 \ \mathrm{and}\  r_{gg}\cdot (r_{b\bar{b}})^{-1}\sim 1,\label{scenario1}\\   
\mathrm{Small}\ \Gamma(H\to b\bar{b})\ \mathrm{scenario}:&\  (r_{{b\bar{b}}})^{-1} > 1\ \mathrm{and}\ r_{gg}\cdot r_{\gamma\gamma}\sim 1. \label{scenario2}
\end{align}
\end{subequations}
Since the two scenarios, light stau scenario and small $\Gamma(H\to b\bar{b})$ scenario, have distinct predictions for $R_{VV}$ and $R_{\tau\bar{\tau}}$, which can be tested in the current run of the LHC, we explore their consequences carefully in the extended parameter space of the MSSM.

For definiteness, we explore the following MSSM parameter region,
\begin{subequations}\label{para}
\begin{align}
5 \le \tan{\beta} \le 40,\ \  &110 \le M_{H^{\pm}} \le 210,\nonumber\\
500\ \mathrm{GeV} \le A_t \le 5000\ \mathrm{GeV},&\ \ 500\ \mathrm{GeV} \le \mu \le 1500\ \mathrm{GeV},\nonumber \\
300\ \mathrm{GeV} \le M_{\tilde{Q}}=&M_{\tilde{U}}=M_{\tilde{D}} \le 1500\ \mathrm{GeV}, \label{para1}
\end{align}
where $M_{H^{\pm}}$ is the charged Higgs boson mass, $M_{\tilde{f}}$ is the SUSY breaking sfermion mass parameter.
The following parameters are set to fixed values,
\begin{align}
A_b=A_{\tau}&=1\ \mathrm{TeV},\nonumber \\
M_3=800\ \mathrm{GeV},\ M_2=200\ &\mathrm{GeV},\ M_1=100\ \mathrm{GeV},\label{para2}
\end{align}
where $M_i$ are gaugino mass parameters, since they do not affect significantly the property of the Higgs bosons. The slepton soft mass parameters are explored in the region
\begin{align}
\ \ 50\ \mathrm{GeV} \le M_{\tilde{L}}=M_{\tilde{E}} \le 500\ \mathrm{GeV} \label{para3}
\end{align}
for the light stau scenario, while it is set to a fixed value
\begin{align}
M_{\tilde{L}}=M_{\tilde{E}}=1\ \mathrm{TeV} \label{para4}
\end{align}
\end{subequations}
for the small $\Gamma(H\to b\bar{b})$ scenario.

\subsection{Fermion and sfermion contributions to $r_{gg}$ and $r_{\gamma\gamma}$}

In the SM, top quark loop contributes dominantly to the $H_{SM}\to gg$ amplitude. The bottom quark loop interferes destructively with the top quark loop for $M_{H_{SM}} \gtrsim 30$ GeV, and for $M_{H_{SM}} \sim 125$ GeV it counteracts the top quark loop contribution by roughly $10$ \%. In the MSSM, the heavier CP even Higgs boson, $H$, couples up and down type fermions, respectively, with the couplings
\begin{subequations}\label{Hff}
\begin{align}
g^{}_{Huu}=\frac{\sin\alpha}{\sin\beta}\left(\frac{\sqrt{2}m_u}{v}\right),\label{Huu}\\
g^{}_{Hdd}=\frac{\cos\alpha}{\cos\beta}\left(\frac{\sqrt{2}m_d}{v}\right).\label{Hdd}
\end{align}
\end{subequations}
Hence, when $\sin\alpha > 0$ in eq. (\ref{+sina}) and $M_{H} \gtrsim 30$ GeV are satisfied, the bottom quark loop interferes destructively with the top quark loop as in the SM. When $\sin\alpha < 0$ in eq. (\ref{-sina}) and $M_{H} \gtrsim 30$ GeV are satisfied, on the other hand, the bottom quark loop interferes constructively with the top quark loop, which can lead to $r_{gg} > 1$.

Similar discussion is applied for $r_{\gamma\gamma}$. In the SM, the W boson loop contributes dominantly to the $H_{SM}\to \gamma\gamma$ amplitude. The sub-dominant top quark loop interferes destructively with the W boson loop, whereas the bottom quark loop interferes constructively, for $M_{H_{SM}} \gtrsim 30$ GeV. In the MSSM, the $H$ coupling to the weak bosons normalized to the SM value is $\cos(\alpha-\beta)$. Hence, when $H$ ($\sim 125$ GeV) has the SM-like coupling to the weak bosons and $\tan\beta \gg 1$, we have $\cos(\alpha-\beta)\simeq \sin\alpha$ and the top quark loop always interferes destructively with the W boson loop, whereas the bottom quark loop interferes constructively when $\sin\alpha > 0$ as in the SM and destructively when $\sin\alpha < 0$.

\begin{figure}[t]
\centering
\includegraphics[scale=0.7]{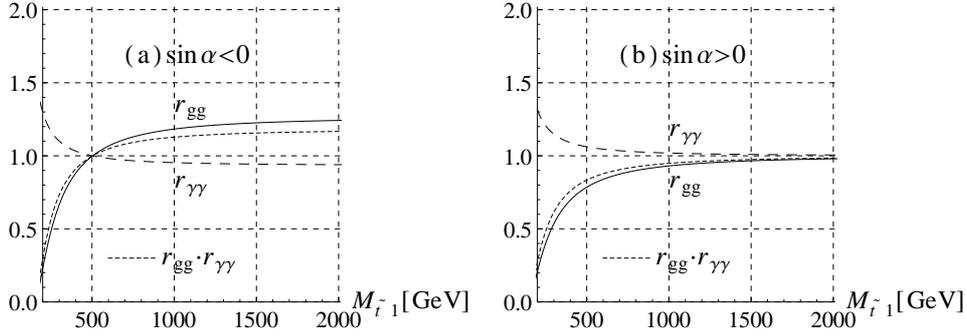}
\caption{$r_{gg}$ (solid line), $r_{\gamma\gamma}$ (dashed line) and $r_{gg}\cdot r_{\gamma\gamma}$ (dotted line) as functions of the lighter stop mass for $\sin\alpha < 0$ (a) and $\sin\alpha > 0$ (b). $M_H=125$ GeV, $A_t/M_{\tilde{Q}}=2.6$, $M_{\tilde{Q}}=M_{\tilde{U}}=M_{\tilde{D}}$, $\tan\beta=10$, $\mu=1$ TeV, $|\sin\alpha|=\sin\beta$ and only the stops are considered among SUSY particles in the amplitudes.}
\end{figure}

Sfermions in the third generation can have important contribution to the $H\to gg$ and the $H\to \gamma\gamma$ amplitudes due to their large Yukawa couplings. The mass eigenstates of the sfermions $\tilde{f}_{1, 2}$ ($M_{\tilde{f}_1} < M_{\tilde{f}_2}$) are mixed states of the current eigenstates $\tilde{f}_{L, R}$ with a mixing angle $\theta_f$,
\begin{align}
\begin{pmatrix}
\tilde{f}_{1}\\
\tilde{f}_{2}
\end{pmatrix}
=
\begin{pmatrix}
\cos\theta_f & -\sin\theta_f\\
\sin\theta_f & \cos\theta_f
\end{pmatrix}
\begin{pmatrix}
\tilde{f}_{L}\\
\tilde{f}_{R}
\end{pmatrix}.\label{thetaf}
\end{align}
The mass matrix of the sfermions in the current basis is given by \cite{masscoupling1, masscoupling2, masscoupling3}
\begin{align}
\begin{pmatrix}
M_{LL}^2 & M_{LR}^2\\
M_{LR}^2 & M_{RR}^2
\end{pmatrix}
=
\begin{pmatrix}
M_{\tilde{f}_L}^2 + m_f^2 + D_L^f & m_f(A_f-\mu r_f)\\
m_f(A_f-\mu r_f) & M_{\tilde{f}_R}^2 + m_f^2 + D_R^f
\end{pmatrix}
, \label{sfmassmatrix}
\end{align}
where $m_f$ is the corresponding fermion mass and $r_d = 1/r_u = \tan\beta$ for down and up type fermions. The $D$ terms are given in terms of the electric charge $e_f$, the weak isospin $I^3_f$ and the weak mixing angle $\theta_w$ by 
\begin{subequations}
\begin{align}
D_L^f&=(I^3_f-e_f\sin^2\theta_w)M_Z^2\cos2\beta, \\
D_R^f&=e_f\sin^2\theta_wM_Z^2\cos2\beta.
\end{align}
\end{subequations}
The mass eigenvalues are
\begin{align}
M^2_{\tilde{f}_{\pm}}=\frac{M_{LL}^2+M_{RR}^2}{2}\pm \frac{1}{2}\sqrt{(M_{LL}^2-M_{RR}^2)^2+4(M_{LR}^2)^2},
\end{align}
where $\tilde{f}_-=\tilde{f}_1$ and $\tilde{f}_+=\tilde{f}_2$ are the lighter and heavier mass eigenstates, respectively, and the mixing angle $\theta_f$ ($|\theta_f| < \pi/2$) is given by
\begin{align}
\sin2\theta_f&=\frac{2m_f(A_f-\mu r_f)}{M_{\tilde{f}_2}^2-M_{\tilde{f}_1}^2},\label{sinthetaf}\\
\cos2\theta_f&=\frac{M_{\tilde{f}_R}^2+D_R^f-M_{\tilde{f}_L}^2-D_L^f}{M_{\tilde{f}_2}^2-M_{\tilde{f}_1}^2}.\label{costhetaf}
\end{align}
The heavier CP even Higgs boson, $H$, couples to up and down type sfermions in the mass eigenstate basis as follows \cite{masscoupling1, masscoupling2, masscoupling3}
\begin{subequations}
\begin{align}
g^{}_{H\tilde{u}_{\pm}\tilde{u}_{\pm}}=\frac{2}{v}\left[m_f^2\frac{\sin\alpha}{\sin\beta}+M_Z^2\cos(\alpha+\beta)\left(I_f^3\cos^2\theta_f-e_f\cos2\theta_f\sin^2\theta_w\right)\right] \nonumber \\
\mp \frac{m_f}{v\sin\beta}\Bigl[\cos\alpha \mu- \sin\alpha A_f\Bigr]\sin2\theta_f, \label{gHuu1}\\
g^{}_{H\tilde{d}_{\pm}\tilde{d}_{\pm}}=\frac{2}{v}\left[m_f^2\frac{\cos\alpha}{\cos\beta}+M_Z^2\cos(\alpha+\beta)\left(I_f^3\cos^2\theta_f-e_f\cos2\theta_f\sin^2\theta_w\right)\right] \nonumber \\
\mp \frac{m_f}{v\cos\beta}\Bigl[\sin\alpha \mu- \cos\alpha A_f\Bigr]\sin2\theta_f. \label{gHdd1}
\end{align}
\end{subequations}
When $H$ is a SM-like Higgs boson with $\tan\beta \gg 1$ and when the mixing between $\tilde{f}_L$ and $\tilde{f}_R$ is large, these couplings are approximated by
\begin{subequations}
\begin{align}
g^{}_{H\tilde{u}_{\pm}\tilde{u}_{\pm}}=\pm\frac{2m_u^2A_u^2}{v\sin\beta(M_{\tilde{u}_2}^2-M_{\tilde{u}_1}^2)}\sin\alpha, \label{gHuu2}\\
g^{}_{H\tilde{d}_{\pm}\tilde{d}_{\pm}}=\pm\frac{2m_d^2\mu^2\tan\beta}{v\cos\beta(M_{\tilde{d}_2}^2-M_{\tilde{d}_1}^2)}\sin\alpha, \label{gHdd2}
\end{align}
\end{subequations}
which are proportional to $\sin\alpha$, and the lighter of the mass eigenstates of the sfermions always interferes destructively with the top quark loop, while the heavier interferes constructively with the top quark loop, independently of the sign of $\sin\alpha$. The lighter one generally contributes dominantly, and hence the squarks with large mixing always reduce $r_{gg}$ and increase $r_{\gamma\gamma}$ at the same time. Figure 1 shows $r_{gg}$ (solid line), $r_{\gamma\gamma}$ (dashed line) and $r_{gg}\cdot r_{\gamma\gamma}$ (dotted line) as functions of the lighter stop mass for $\sin\alpha < 0$ (left) and $\sin\alpha > 0$ (right), where $M_H=125$ GeV, $M_{\tilde{Q}}=M_{\tilde{U}}=M_{\tilde{D}}$, $A_t=2.6M_{\tilde{Q}}$, $\tan\beta=10$, $\mu=1$ TeV, $|\sin\alpha|=\sin\beta$ and only the stops are considered among SUSY particles in the amplitudes. In both cases, the reduction of $r_{gg}$ due to stop contribution is always larger than the corresponding enhancement in $r_{\gamma\gamma}$, and hence the light stop reduces the product $r_{gg}\cdot r_{\gamma\gamma}$. When $\tan\alpha = -\tan\beta = -10$, the bottom quark contributes constructively to the top quark loop, giving $r_{gg} > 1$ in Figure 1 (a) for large stop masses. In contrast to squarks, stau can increase $r_{\gamma\gamma}$ without affecting $r_{gg}$.

\subsection{Light stau scenario}
In this section, we examine $R_{\gamma\gamma}$, $R_{VV}$ and $R_{\tau\bar{\tau}}$ in the light stau scenario of eq. (\ref{scenario1}) where the mass of the heavier CP even Higgs boson is $125\pm2$ GeV and a large $R_{\gamma\gamma}$ is obtained by increasing $r_{\gamma\gamma}$ .

\subsubsection{$R_{\gamma\gamma}$}

As discussed in Section 3.1, light stau with large mixing between $\tilde{\tau}_L$ and $\tilde{\tau}_R$ can increase $r_{\gamma\gamma}$ without decreasing $r_{gg}$. Hence, with light stau and heavy squarks, we can expect
\begin{align}
r_{gg}\cdot r_{\gamma\gamma} > 1. \label{something2}
\end{align}
In Figure 2 (left), we show the maximum and minimum values of $R_{\gamma\gamma}$ as functions of the mass difference between the lighter and heavier staus, for three different masses of the lighter stau, 82 GeV (solid line), 100 GeV (dotted line) and 140 GeV (dashed line). We impose in the light stau scenario the condition
\begin{align}
0.9 < r_{b\bar{b}} < 1.1,\label{Gbb}
\end{align}
so that the enhancement of $R_{\gamma\gamma}$ is mostly due to the light stau contribution to $r_{\gamma\gamma}$. The plot shows that light stau increases $R_{\gamma\gamma}$ as the mass difference grows, as expected, since large mass difference corresponds to large mixing between $\tilde{\tau}_L$ and $\tilde{\tau}_R$. The maximum value of $R_{\gamma\gamma}$ is obtained when the lighter stop has large mass $\simeq 1300$ GeV and the stau mixing is large, while the minimum value of $R_{\gamma\gamma}$ is obtained when the lighter stop has small mass $\simeq 300$ GeV and the stau mixing is small, within our explored parameter region of eq. (\ref{para}), since light squarks generally decrease $r_{gg}\cdot r_{\gamma\gamma}$, as discussed in Section 3.1.

\begin{figure}[t]
\centering
\includegraphics[scale=0.8]{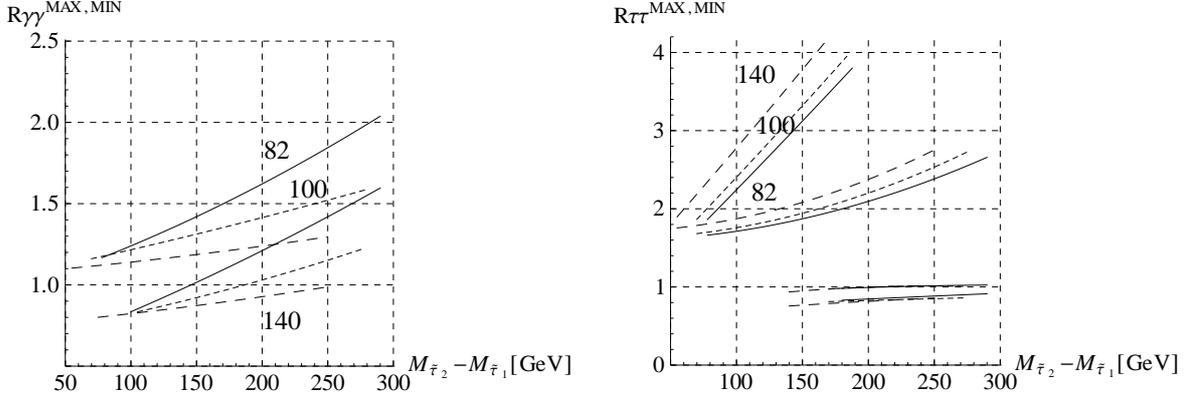}
\caption{The maximum and minimum values of $R_{\gamma\gamma}$ (left) and $R_{\tau\bar{\tau}}$ (right) are plotted against the mass difference between the lighter and heavier staus, for three different masses of the lighter stau, 82 GeV (solid line), 100 GeV (dotted line) and 140 GeV (dashed line) in the light stau scenario. It is shown that light stau increases $R_{\gamma\gamma}$ as the mass difference grows. The maximum value of $R_{\gamma\gamma}$ is obtained when the lighter stop has large mass $\simeq 1300$ GeV and the stau mixing is large, while the minimum value of $R_{\gamma\gamma}$ is obtained when the lighter stop has small mass $\simeq 300$ GeV and the stau mixing is small, within our explored parameter region of eq. (\ref{para}), since light squarks generally decrease $r_{gg}\cdot r_{\gamma\gamma}$, as discussed in Section 3.1. The enhanced $R_{\tau\bar{\tau}}$ appears when $\sin\alpha < 0$, while $R_{\tau\bar{\tau}}\sim 1$ is obtained when $\sin\alpha > 0$, see Section 3.2.3.}
\end{figure}

\subsubsection{$R_{VV}$}

Next we consider $R_{VV}\ (V=W,\ Z)$ in the light stau scenario. Even with the help of the light stau contribution, significant enhancement of $R_{\gamma\gamma}$ over unity is possible only when the heavier CP even Higgs boson $H$ has the SM-like coupling to the weak bosons as explained in Section 2. We therefore expect $R_{VV} \sim r_{VV} \sim 1$ in this scenario, although $r_{VV} < 1$ always holds. In our numerical calculation, $r_{VV}$ lies between $0.95$ and $1$, whereas $R_{VV}$ is found to lie between $0.7$ and $1.15$ for the lighter stop mass between $300$ and $1300$ GeV, with little dependence on the stau masses. $R_{VV} > 1.1$ is found when $\sin\alpha < 0$ where the bottom quark contributes constructively with the top quark loop, giving $r_{gg} > 1.1$, see Figure 1 (a).

\subsubsection{$R_{\tau\bar{\tau}}$}
Here we discuss $R_{\tau\bar{\tau}}$ in the light stau scenario. As discussed above, the enhancement of $R_{\gamma\gamma}$ by the light staus can be obtained for large values of $\mu$ and $\tan\beta$ which appears in the $H\tilde{\tau}_i\tilde{\tau}_i$ ($i=-,+$ or $1,2$) couplings in eq. (\ref{gHdd2}). When $\mu\tan{\beta}$ is large, the radiative SUSY corrections in the bottom quark and the tau lepton masses can be important,
\begin{subequations}
\begin{align}
M_b=\frac{Y_b}{\sqrt{2}}v\cos{\beta}(1+\Delta_b),\\
M_{\tau}=\frac{Y_{\tau}}{\sqrt{2}}v\cos{\beta}(1+\Delta_{\tau}),
\end{align}
\end{subequations}
where \cite{deltab1, deltab2, deltab3}
\begin{subequations}
\begin{align}
\Delta_b&=\mu\tan\beta\left[\frac{2\alpha_s}{3\pi}M_{3} I(M_{\tilde{b}_1},M_{\tilde{b}_2},M_{3})+\frac{Y_t^2}{16\pi^2}A_t I(M_{\tilde{t}_1},M_{\tilde{t}_2},\mu)\right], \label{Deltab}\\
\Delta_{\tau}&=\mu\tan\beta \left[\frac{g_1^2}{16\pi^2}M_{1} I(M_{\tilde{\tau}_1},M_{\tilde{\tau}_2},M_{1})+\frac{g_2^2}{16\pi^2}M_2 I(M_{\tilde{\nu}_{\tau}},M_{2},\mu)\right]. \label{Deltatau}
\end{align}
\end{subequations}
The function $I(a,b,c)$ is given by \cite{deltab1, deltab2, deltab3}
\begin{align}
I(a,b,c)=\frac{a^2b^2\ln(a^2/b^2)+b^2c^2\ln(b^2/c^2)+c^2a^2\ln(c^2/a^2)}{(a^2-b^2)(b^2-c^2)(a^2-c^2)},
\end{align}
which is positive for all real $a, b, c$.
The effective Higgs couplings to $b\bar{b}$ and $\tau\bar{\tau}$ normalized to the SM values are now given by \cite{hbb1, hbb2} 
\begin{subequations}
\begin{align}
g^{}_{Hb\bar{b}}=\frac{\cos\alpha}{\cos\beta}\left[1-\frac{\Delta_b}{1+\Delta_b}\left(1-\frac{\tan\alpha}{\tan\beta}\right)\right], \label{gHbb}\\
g^{}_{H\tau\bar{\tau}}=\frac{\cos\alpha}{\cos\beta}\left[1-\frac{\Delta_{\tau}}{1+\Delta_{\tau}}\left(1-\frac{\tan\alpha}{\tan\beta}\right)\right], \label{gHtautau}
\end{align}
\end{subequations}
and their squared values should approximately correspond to $r_{b\bar{b}}$ and $r_{\tau\bar{\tau}}$, respectively. The behavior of these couplings strongly depends on the mixing angle $\alpha$, $-\frac{\pi}{2} < \alpha < 0$ ($\sin\alpha < 0$) in eq. (\ref{-sina}) or $0 < \alpha < \frac{\pi}{2}$ ($\sin\alpha > 0$) in eq. (\ref{+sina}). Figure 2 (right) shows the maximum and minimum values of $R_{\tau\bar{\tau}}$ against the mass difference between the lighter and heavier staus, for three different masses of the lighter stau, 82 GeV (solid line), 100 GeV (dotted line) and 140 GeV (dashed line) in the light stau scenario. The enhanced $R_{\tau\bar{\tau}}$ in the plot appears when $\sin\alpha < 0$ and it lies roughly between $2$ and $4$ even when $r_{b\bar{b}} \sim 1$, whereas $R_{\tau\bar{\tau}}\sim 1$ appears when $\sin\alpha > 0$. Below we discuss $R_{\tau\bar{\tau}}$ in detail for each case.\\

\underline{$R_{\tau\bar{\tau}}$ when $\sin\alpha < 0$}\\
Since $H$ is now the SM-like Higgs boson with $\cos(\alpha-\beta) \simeq -1$ in eq. (\ref{H}), we can estimate its deviation from the SM limit, with a small parameter $\epsilon$, as
\begin{align}
\frac{-\tan\alpha}{\tan\beta}=1-\epsilon.
\end{align}
The effective Higgs coupling to $b\bar{b}$ in eq. (\ref{gHbb}) can then be expressed as
\begin{align}
g_{Hb\bar{b}}= \frac{1}{1-\epsilon}\left[1-\frac{\Delta_b}{1+\Delta_b}(2-\epsilon)\right],\label{eq2}
\end{align}
while the coupling to $\tau\bar{\tau}$ in eq. (\ref{gHtautau}) may be approximated as
\begin{align}
g_{H\tau\bar{\tau}} = \frac{1}{1-\epsilon},
\end{align}
since $\Delta_{\tau}$ is significantly smaller than $\Delta_b$ due to the electroweak couplings in eq. (\ref{Deltatau}).
With the above approximation, the partial width ratio $(r_{b\bar{b}})^{-1}\cdot r_{\tau\bar{\tau}}$ in eq. (\ref{Rtautau}) is calculated and we find
\begin{align}
R_{\tau\bar{\tau}}\simeq r_{gg} \cdot \left(\frac{1+\Delta_b}{1-\Delta_b(1-\epsilon)}\right)^2.\label{eq31}
\end{align}
$R_{\tau\bar{\tau}}$ should hence be always larger than unity if $\Delta_b$ is positive and $r_{gg}\sim1$. We note that $\Delta_b$ is positive in the MSSM parameter region of eq. (\ref{para}) we explore in this study, where $\mu$, $M_3$ and $A_t$ are all positive.

Our assumption of $r_{b\bar{b}}\approx 1$ in the light stau scenario of eq. (\ref{Gbb}) can be satisfied when $\epsilon\approx 2\Delta_b/(1+2\Delta_b)$,
which leads to 
\begin{align}
R_{\tau\bar{\tau}} \simeq r_{gg}\cdot \left(1+2\Delta_b\right)^2.\label{eq1}
\end{align}
The increase of $R_{\tau\bar{\tau}}$ for large mixing between $\tilde{\tau}_L$ and $\tilde{\tau}_R$ found for $\sin\alpha < 0$ in Figure 2 (right) can be explained by this mechanism, since the large mass splitting $M_{\tilde{\tau}_2}-M_{\tilde{\tau}_1}$ implies large $\mu\tan\beta$ in eq. (\ref{sfmassmatrix}), which leads to large $\Delta_b$ from eq. (\ref{Deltab}). In contrast to $R_{\gamma\gamma}$, large values of $R_{\tau\bar{\tau}}$ are found for light squarks, because increase of $\Delta_b$ induced by light squarks in eq. (\ref{Deltab}) is larger than the decrease of $r_{gg}$. The maximum value of $R_{\tau\bar{\tau}}$ is obtained when mass of the lighter stop is roughly between $400$ and $600$ GeV and the stau mixing is large, while the minimum value of $R_{\tau\bar{\tau}}$ is obtained when mass of the lighter stop is $1300$ GeV and the stau mixing is small, within our explored parameter region of eq. (\ref{para}).\\

\underline{$R_{\tau\bar{\tau}}$ when $\sin\alpha > 0$}\\
In this case, we can express $\tan\alpha/\tan\beta$ $(>0)$ as
\begin{align}
\frac{\tan\alpha}{\tan\beta}=1-\epsilon.
\end{align}
The effective Higgs coupling to $b\bar{b}$ in eq. (\ref{gHbb}) now becomes
\begin{align}
g_{Hb\bar{b}}= \frac{1}{1-\epsilon}\left[1-\frac{\Delta_b}{1+\Delta_b}\epsilon\right],\label{eq5}
\end{align}
while the coupling to $\tau\bar{\tau}$ in eq. (\ref{gHtautau}) may be approximated again as
\begin{align}
g_{H\tau\bar{\tau}} = \frac{1}{1-\epsilon}.
\end{align}
We then find 
\begin{align}
R_{\tau\bar{\tau}}\simeq r_{gg} \cdot \left(\frac{1+\Delta_b}{1+\Delta_b(1-\epsilon)}\right)^2.\label{eq41}
\end{align}

Our assumption of $r_{b\bar{b}}\approx 1$ in the light stau scenario of eq. (\ref{Gbb}) can be satisfied only when $|\epsilon| \lesssim 0.1$, which implies
\begin{align}
R_{\tau\bar{\tau}}\simeq r_{gg}.\label{eq51}
\end{align}
This behavior of $R_{\tau\bar{\tau}}$ is shown in Figure 2 (right). The plot shows little dependence of $R_{\tau\bar{\tau}}$ on the stau masses and mixing as it is expected from eq. (\ref{eq51}) when $\sin\alpha > 0$. The maximum values of $R_{\tau\bar{\tau}}$ in the plot are obtained when mass of the lighter stop is around $1300$ GeV, while the minimum values of $R_{\tau\bar{\tau}}$ are obtained when mass of the lighter stop is between $300$ and $400$ GeV within our explored parameter region, since light squarks decrease $r_{gg}$, as discussed in Section 3.1.

Summing up this sub-section, the sub-scenarios with $\sin\alpha\ (\sim \sin\beta) > 0$ and $\sin\alpha\ (\sim -\sin\beta) < 0$ in eqs. (\ref{sina}) can be distinguished in the light stau scenario by measuring $R_{\tau\bar{\tau}}$. If no significant enhancement over the SM rate is found, only the $\sin\alpha > 0$ region is allowed, where the loop contribution reverses the sign of the off-diagonal element $M_{ud}^2$ of the Higgs mass squared matrix in eq. (\ref{Mud}).

\begin{figure}[t]
\centering
\includegraphics[scale=0.65]{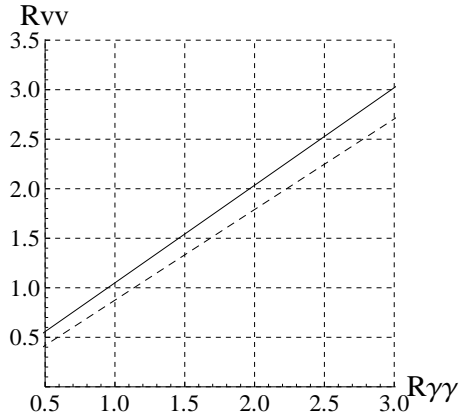}
\caption{Correlation between $R_{\gamma\gamma}$ and $R_{VV}$ with the maximum (solid line) and the minimum (dashed line) values of $R_{VV}$ in the small $\Gamma(H\to b\bar{b})$ scenario. At some point of $R_{\gamma\gamma}$, the maximum value of $R_{VV}$ is obtained for heavy squarks, while the minimum value is obtained for light squarks.}
\end{figure}

\subsection{Small $\Gamma(H\to b\bar{b})$ scenario}
If the stau masses are as large as 1 TeV, their contribution to $r_{\gamma\gamma}$ is suppressed, and the only alternative way to enhance $R_{\gamma\gamma}$ is to suppress $\Gamma(H\to b\bar{b})$ in eq. (\ref{Rpp}),
\begin{align}
(r_{b\bar{b}})^{-1} > 1. \label{condition11}
\end{align}
In this section, we examine $R_{\gamma\gamma}$, $R_{VV}$ and $R_{\tau\bar{\tau}}$ in the small $\Gamma(H\to b\bar{b})$ scenario of eq. (\ref{scenario2}) where the mass of the heavier CP even Higgs boson is $125\pm2$ GeV and an enhanced value of $R_{\gamma\gamma}$ is obtained by decreasing $r_{b\bar{b}}$ .

\subsubsection{$R_{\gamma\gamma}$ and $R_{VV}$}
When $r_{b\bar{b}}$ is suppressed, not only $R_{\gamma\gamma}$ but also $R_{VV}$ should be enhanced. Figure 3 shows the correlation between $R_{\gamma\gamma}$ and $R_{VV}$, with the maximum (solid line) and the minimum (dashed line) values of $R_{VV}$ for a given $R_{\gamma\gamma}$ in the small $\Gamma(H\to b\bar{b})$ scenario. When we compare the maximum and minimum values of $R_{VV}$ for a given $R_{\gamma\gamma}$, the maximum value is obtained for heavy squarks, while the minimum value is obtained for light squarks. This is because light stop contributions enhance $r_{\gamma\gamma}$ whereas their contributions to $r_{gg}$ are common in $R_{\gamma\gamma}$ and $R_{VV}$. We find that both $R_{\gamma\gamma}$ and $R_{VV}$ can be as large as $3$ in the small $\Gamma(H\to b\bar{b})$ scenario within our explored parameter region of eq. (\ref{para}).

\subsubsection{$R_{\tau\bar{\tau}}$}

As in the light stau scenario, $R_{\tau\bar{\tau}}$ in the small $\Gamma(H\to b\bar{b})$ scenario depends strongly on the sign of $\sin\alpha$. We show in Figure 4 the maximum and minimum values of $R_{\tau\bar{\tau}}$ as functions of the product of $\mu$ and $\tan\beta$, for three different masses of the lighter stop, 400 GeV (solid line), 600 GeV (dotted line) and 1000 GeV (dashed line), when $R_{\gamma\gamma}$ lies between $1.9$ and $2.1$.

\begin{figure}[t]
\centering
\includegraphics[scale=0.9]{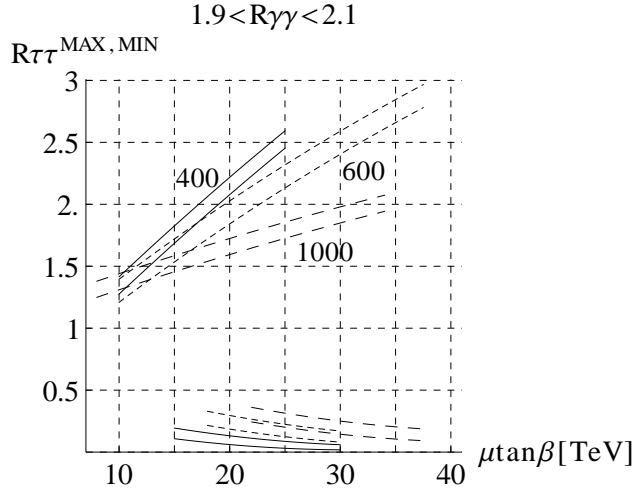}
\caption{The maximum and minimum values of $R_{\tau\bar{\tau}}$ as functions of the product of $\mu$ and $\tan\beta$, for three different masses of the lighter stop, 400 GeV (solid line), 600 GeV (dotted line) and 1000 GeV (dashed line) in the small $\Gamma(H\to b\bar{b})$ scenario, when $R_{\gamma\gamma}$ lies between $1.9$ and $2.1$. The enhanced $R_{\tau\bar{\tau}}$ appears when $\sin\alpha < 0$, while suppressed $R_{\tau\bar{\tau}} < 0.5$ is obtained when $\sin\alpha > 0$, see Section 3.3.2.}
\end{figure}

The enhanced $R_{\tau\bar{\tau}}$ in Figure 4 appears when $\sin\alpha < 0$. The increase of $R_{\tau\bar{\tau}}$ with a rise in $\mu\tan\beta$ is explained from eqs. (\ref{eq31}, \ref{Deltab}), since $R_{\tau\bar{\tau}}$ increases with $\Delta_b$ in eq. (\ref{eq31}) and $\Delta_b$ increases with $\mu\tan\beta$ in eq. (\ref{Deltab}). 
Larger values of $R_{\tau\bar{\tau}}$ can be obtained for lighter stops because the reduction in $r_{b\bar{b}}$ due to the $\Delta_b$ contribution is bigger than the reduction in $r_{gg}$, as discussed for the light stau scenario below eq. (\ref{eq1}).

The highly suppressed $R_{\tau\bar{\tau}}$ in Figure 4 appears when $\sin\alpha > 0$. From eq. (\ref{eq5}), $r_{b\bar{b}} < 1$ can be 
obtained when $\epsilon < 0$, or $\tan\alpha > \tan\beta$. 
The plot shows that $R_{\tau\bar{\tau}}$ decreases with a rise in $\mu\tan\beta$ in contrast to the case of $\sin\alpha < 0$. This is because $R_{\tau\bar{\tau}}$ in eq. (\ref{eq41}) decreases as $\Delta_b$ grows when $\epsilon < 0$. The suppression of $R_{\tau\bar{\tau}}$ for light stops can be explained as follows. First, light stops lead to small $r_{gg}$, as discussed in Section 3.1. Secondly, since light squarks give small $r_{gg}$, $r_{b\bar{b}}$ must get smaller in order to keep $R_{\gamma\gamma}$ between $1.9$ and $2.1$. This requires even smaller $\epsilon\ ( < 0)$ in eq. (\ref{eq5}), which further suppresses $R_{\tau\bar{\tau}}$ as in eq. (\ref{eq41}).

\section{Constraints on the other Higgs bosons}

In contrast to the decoupling scenarios of the MSSM, in our two scenarios where the heavier of the CP even Higgs bosons has mass around 125 GeV, and at the same time, has the SM-like (nearly maximum) coupling to the weak bosons,
none of the other Higgs bosons can be very heavy. The masses of the CP odd and charged Higgs bosons are bounded from above when $H$ is a SM-like Higgs boson, as explained in Section 2; see eqs. (\ref{eq81}, \ref{eq42}). From eq. (\ref{eq42}), the largest $M_A$ may be obtained when
\begin{align}
\bar{A_t}^2=6+\frac{\bar{\mu}^2}{2},\label{mamax}
\end{align}
whereas in order to obtain $M_H$ as large as 125 GeV, $\bar{A_t}^2\sim 6$ is necessary from eqs. (\ref{Muu}, \ref{mH3}). 

Figure 5 shows the allowed mass regions in the $M_h$ and $M_A$ space (lower regions) and in the $M_{H^{\pm}}$ and $M_A$ space (upper regions) when $1.5 < R_{\gamma\gamma} < 2.5$, for the light stau scenario (dashed line) and the small $\Gamma(H\to b\bar{b})$ scenario (solid line). The mass region with large $M_A$ and $M_H^{\pm}$ are obtained by the large radiative SUSY correction to the Higgs potential in eq. (\ref{eq42}). The lower bound on $M_h$ comes from the upper bound on the cross section $\sigma(e^+e^-\to Zh)$, and the lower bound on $M_h+M_A$ comes from the upper bound on the cross section $\sigma(e^+e^-\to Ah)$. The main reason for the smallness of the allowed regions in the light stau scenario is because $r_{b\bar{b}}$ is constrained to be between $0.9$ and $1.1$ in the light stau scenario, while it is not constrained in the small $\Gamma(H\to b\bar{b})$ scenario. Since $h$ is non SM-like Higgs boson in our scenarios, the cross section $\sigma(e^+e^-\to Zh)$ is highly suppressed, while the cross sections $\sigma(e^+e^-\to ZH)$ and  $\sigma(e^+e^-\to Ah)$ are not suppressed. Hence all these Higgs bosons should be discovered in the future $e^+e^-$ collider. Productions of $h$ and $A$ bosons at the LHC are dominated by gluon fusion via the bottom quark loop or bottom quark annihilation, because of their suppressed couplings to the weak bosons and the top quark. They can be discovered in the $\tau^+\tau^-$ decay channel, especially at large $\tan\beta$. The charged Higgs boson with $M_{H^{\pm}} \lesssim 150$ GeV may be discovered from the top quark decay at the LHC.

\begin{figure}[t]
\centering
\includegraphics[scale=0.75]{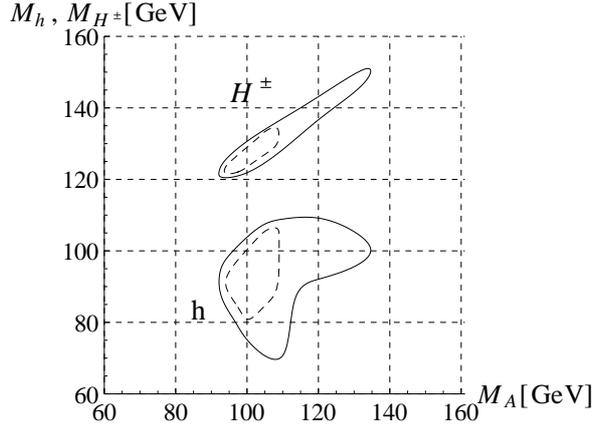}
\caption{Allowed mass regions in the $M_h$ and $M_A$ space (lower plots) and in the $M_{H^{\pm}}$ and $M_A$ space (upper plots) when $1.5 < R_{\gamma\gamma} < 2.5$, for the light stau scenario (dashed line) and the small $\Gamma(H\to b\bar{b})$ scenario (solid line).}
\end{figure}

\section{Conclusion}
In this study, we identify the 125 GeV state as the heavier of CP even Higgs bosons in the MSSM, and study two scenarios where $R_{\gamma\gamma}$, the two photon production rate normalized to the SM prediction, can be significantly larger than unity.

In one scenario, the $H \to \gamma\gamma$ amplitude is enhanced by the light stau contribution which interferes constructively with the main W boson loop contribution. Within our explored parameter region, we find that $R_{\gamma\gamma}$ as large as 2.0 can be obtained with a light stau near the current mass bound ($M_{\tilde{\tau}_1} = 82$ GeV), when the mixing between $\tilde{\tau}_L$ and $\tilde{\tau}_R$ proportional to $\mu\tan\beta$ is large and when the squarks are heavy. The $WW^* (ZZ^*)$ rate, $R_{VV}$, has little dependence on the stau masses and mixing, and we find that $R_{VV}$ can be between $0.7$ for the lighter stop mass $\approx 300$ GeV and $1.15$ for the lighter stop mass $\approx 1300$ GeV. Due to the large radiative SUSY correction to the bottom quark mass, large $\tau\bar{\tau}$ rate, $R_{\tau\bar{\tau}}$, between 2.0 and 4.0 can be obtained even when $R_{b\bar{b}}$ is around unity. The maximum value of $R_{\tau\bar{\tau}}$ is obtained for the large stau mixing and for the lighter stop mass between $400$ and $600$ GeV, with little dependence on the lighter stau mass. This enhanced $R_{\tau\bar{\tau}}$ appears when the mixing angle $\alpha$ of the CP even Higgs bosons in the basis of the two Higgs doublets is $-\pi/2 < \alpha < 0$, while $R_{\tau\bar{\tau}}$ around unity is also possible when $0 < \alpha < \pi/2$.

In another scenario, $R_{\gamma\gamma}$ is enhanced by suppressing the dominant partial decay width $\Gamma(H\to b\bar{b})$, and not only $R_{\gamma\gamma}$ but also $R_{VV}$ is enhanced. We find that both $R_{\gamma\gamma}$ and $R_{VV}$ can be as large as $3$ within our explored MSSM parameter region. As in the light stau scenario, $R_{\tau\bar{\tau}}$ can be enhanced when $-\pi/2 < \alpha < 0$, while suppressed when $0 < \alpha < \pi/2$. We find that $R_{\tau\bar{\tau}}$ can be as large as $2.5$, for example, for the lighter stop mass $\simeq 400$ GeV and $\mu\tan\beta \simeq 25$ TeV or for the lighter stop mass $\simeq 600$ GeV and $\mu\tan\beta \simeq 30$ TeV when $-\pi/2 < \alpha < 0$, whereas $R_{\tau\bar{\tau}}$ can be as small as $0.1$ with little dependence on the lighter stop mass and on $\mu\tan\beta$ when $0 < \alpha < \pi/2$, even when we assume that $R_{\gamma\gamma}$ lies between $1.9$ and $2.1$.

We also study mass spectra of other three Higgs bosons, $h$, $A$ and $H^{\pm}$ in both scenarios when $R_{\gamma\gamma}$ lies between $1.5$ and $2.5$. We find in both scenarios that all the masses are bounded from above within our explored parameter regions, such that single and pair production of all the Higgs bosons should be observed in $e^+e^-$ collisions at $\sqrt{s} \lesssim 300$ GeV. The charged Higgs boson mass should lie in the region $120 \lesssim M_{H^{\pm}} \lesssim 150$ GeV and may be discovered in the top quark decays at the LHC.

\acknowledgments{J.N. thanks Yoshitaro Takaesu for valuable discussions. J.S.L wishes to thank KEK Theory Center for support and hospitality during his visit, where part of this study has been carried out. The work of J.S.L. is supported in part by the National Science Council of Taiwan under Grant No. 100-2112-M-007-023-MY3. This work is also supported in part by Grant-in-Aid for scientific research (\#20340064 and \#23104006) from Japan Society for the Promotion of Science. We are thankful that the calculation in this study was executed on the Central Computing System of KEK.}

\end{document}